\documentclass[prb,twocolumn, superscriptaddress]{revtex4}
\usepackage[pdftex]{graphicx}
 \usepackage{amsmath}
\usepackage[T1]{fontenc}
\usepackage{flushend}
\usepackage{amssymb}
\usepackage{amsfonts}
\usepackage{bm}
 \usepackage{amsmath} 
\usepackage{lipsum}
\usepackage{amsfonts} 
\usepackage{amssymb, mathrsfs}
\usepackage{braket}
\usepackage{graphicx} 
\usepackage{subfigure}
\usepackage{bbm}
\def\beq{\begin{equation}}
\def\eeq{\end{equation}}
\def\bsp{\begin{split}}
\def\esp{\end{split}}
\def\bea{\begin{eqnarray}}
\def\eea{\end{eqnarray}}
\def\ba{\begin{array}}
\def\ea{\end{array}}

\def\dg{\dagger}

\def\lb{\left(}
\def\rb{\right)}

\def\l.{\left.}
\def\r.{\right.}

\def\ra{\rangle}
\def\la{\langle}

\def\bo{\bold{k}}

\begin{document}

\date{\today}
\title{Topological magnon  bands and unconventional thermal Hall effect on the frustrated honeycomb and bilayer triangular lattice}
\email{sowerre@perimeterinstitute.ca}
\author{S. A. Owerre}
\affiliation{Perimeter Institute for Theoretical Physics, 31 Caroline St. N., Waterloo, Ontario N2L 2Y5, Canada.}

\begin{abstract}
In the conventional ferromagnetic systems, topological magnon bands and thermal Hall effect are due to the Dzyaloshinskii-Moriya  interaction (DMI). In principle, however, the DMI is either negligible or it is not allowed by symmetry in some quantum magnets. Therefore, we expect that topological magnon  features will not be present in those   systems. In addition, quantum magnets on the triangular-lattice are not expected to possess topological features  as the DMI or spin-chirality cancels out due to equal and opposite contributions from adjacent triangles.  Here, however, we predict that the isomorphic frustrated honeycomb-lattice and bilayer triangular-lattice  antiferromagnetic system will exhibit  topological magnon bands and topological thermal Hall effect in the absence of an intrinsic DMI. These unconventional topological magnon features  are present  as a result of magnetic-field-induced  non-coplanar spin configurations with nonzero scalar spin chirality.  The relevance of the results to  realistic bilayer triangular antiferromagnetic materials are discussed.
\end{abstract}
\maketitle

\section{Introduction}

 In recent years, the concept of topological band theory has been extended to bosonic systems. Consequently, topological magnon bands and the associated  thermal Hall effect   in insulating quantum ferromagnets have garnered considerable attention. The experimental realizations of these phenomena in the quasi-two-dimensional (2D) kagom\'e ferromagnets Cu(1-3, bdc)  \cite{alex6,rc} have further rekindled much interest in this area. Thermal Hall effect of magnons  was previously realized  experimentally in different 3D pyrochlore ferromagnets A$_2$B$_2$O$_7$  \cite{alex1,alex1a}. These experimental studies follow from different theoretical proposals \cite{alex0, alex2, zhh, alex4, alex4h,shin1,shin, kov1,rol}. An extension to unfrustrated  honeycomb magnets has been recently proposed by different authors \cite{sol,sol1,kkim,ran,kov, sol3}.

 Generally speaking, it is believed that the topological magnon phenomena  in quantum ferromagnets \cite{alex6,rc,alex1,alex1a} result from the   DMI \cite{dm, dm2}, which plays the same role as spin-orbit coupling (SOC) in electronic systems \cite{fdm, yu3}.   To date, there is no experimental evidence of a counter-example where topological magnon features originate from an alternative source other than  the DMI in any magnetically ordered system. Therefore, the conception is that the DMI is mandatory for topological magnon properties to exist in magnetically ordered systems.

The frustrated magnets provide a platform to explore this possibility as we have previously shown on the kagom\'e-type lattices \cite{sa,sa1}.  But the kagom\'e-type lattices naturally allow an intrinsic DMI,  which is capable of inducing and stabilizing the coplanar spin structure \cite{el}. In contrast, most non-kagom\'e-type  frustrated magnets  do not allow an intrinsic DMI due to symmetry. Therefore,  we wish to extend our analyses to those systems. On the honeycomb lattice, geometric frustration is present when a next-nearest-neighbour (NNN) antiferromagnetic interaction $\mathcal J_2$ competes with a nearest-neighbour (NN) antiferromagnetic interaction  $\mathcal J_1$. This system is known as the frustrated $\mathcal J_1$--$\mathcal J_2$ honeycomb-lattice Heisenberg antiferromagnet (HLHAF). It possesses interesting phase diagram for  $\mathcal J_2/\mathcal J_1\ll 1$ \cite{matt,mak0, mak1, mak2, mak3, mak4,mak6,mak7, mak8, mak9, mak10, mak11, mak12,mak13}. The opposite limit  $\mathcal J_2/\mathcal J_1\gg 1$ is unexplored. In this regime the geometric frustration induced by $\mathcal J_2$  yields a  decoupled $120^\circ$ coplanar order. The model  is now isomorphic to a  $120^\circ$ coplanar order on the stacked bilayer triangular-lattice Heisenberg antiferromagnet (TLHAF) with intraplane coupling $\mathcal J_2$ and small interplane coupling $\mathcal J_1$ \cite{jian}. Therefore, one can capture the physics of bilayer TLHAF by studying the dominant $\mathcal J_2$ limit of the frustrated HLHAF.  

 In this paper, we predict that topological magnon bands and unconventional topological thermal Hall effect will exist on the isomorphic honeycomb and bilayer triangular lattice $\mathcal J_1$--$\mathcal J_2$ model without any DMI.   We show that the system possesses topological magnon bands with nonzero Chern number $\mathcal C_\pm=\pm\text{sgn}(\sin \phi)$, where $\sin\phi$ is related to the field-induced  scalar spin chirality $\chi= \sum \boldsymbol{\mathcal S}_i\cdot\lb \boldsymbol{\mathcal S}_{j}\times\boldsymbol{\mathcal S}_{k}\rb$, and $i,j,k$ label sites on a unit triangle;  $\phi$ is the angle subtended by three non-coplanar (umbrella) spins.   The corresponding thermal Hall conductivity $\kappa_{xy}$ is tunable by the external magnetic field as it requires no DMI. Interestingly,  topological properties are not expected to be present on the triangular lattice as the DMI (spin-chirality) cancels out due to equal and opposite contributions from adjacent triangles \cite{alex0}. The current result is a counter-example where the field-induced scalar-chirality of the  non-coplanar (umbrella) spin structure does not cancel as it is coupled to the magnetization of the non-coplanar spins \cite{gro}.
 
 These results are particularly interesting especially  for the stacked triangular-lattice antiferromagnetic materials with no intrinsic DMI. They include Ba$_3$XSb$_2$O$_9$ (X $\equiv$ Mn, Co, and Ni)   \cite{doi,shi,ma, maksi, shi1, qui, zhou} and VX$_2$  (X $\equiv$ Cl, Br, and I) \cite{hir,hir1} and others \cite{coll}.   The effects of topological magnons are also manifested by the measurement of nonzero  thermal Hall conductivity $\kappa_{xy}$ at various external magnetic fields along the $\hat z$-axis  \cite{alex6,rc}. We note that the quasi-2D bilayer metallic triangular-lattice magnet  PdCrO$_2$ with 120$^\circ$ coplanar order also shows a finite anomalous Hall effect in a perpendicular-to-plane external magnetic field \cite{tak}. We therefore expect that an analogous thermal Hall effect in magnetic insulators with charge-neutral excitations such as magnons will be present and of great importance.


\section{Model}
\begin{figure}
\centering
\includegraphics[width=1\linewidth]{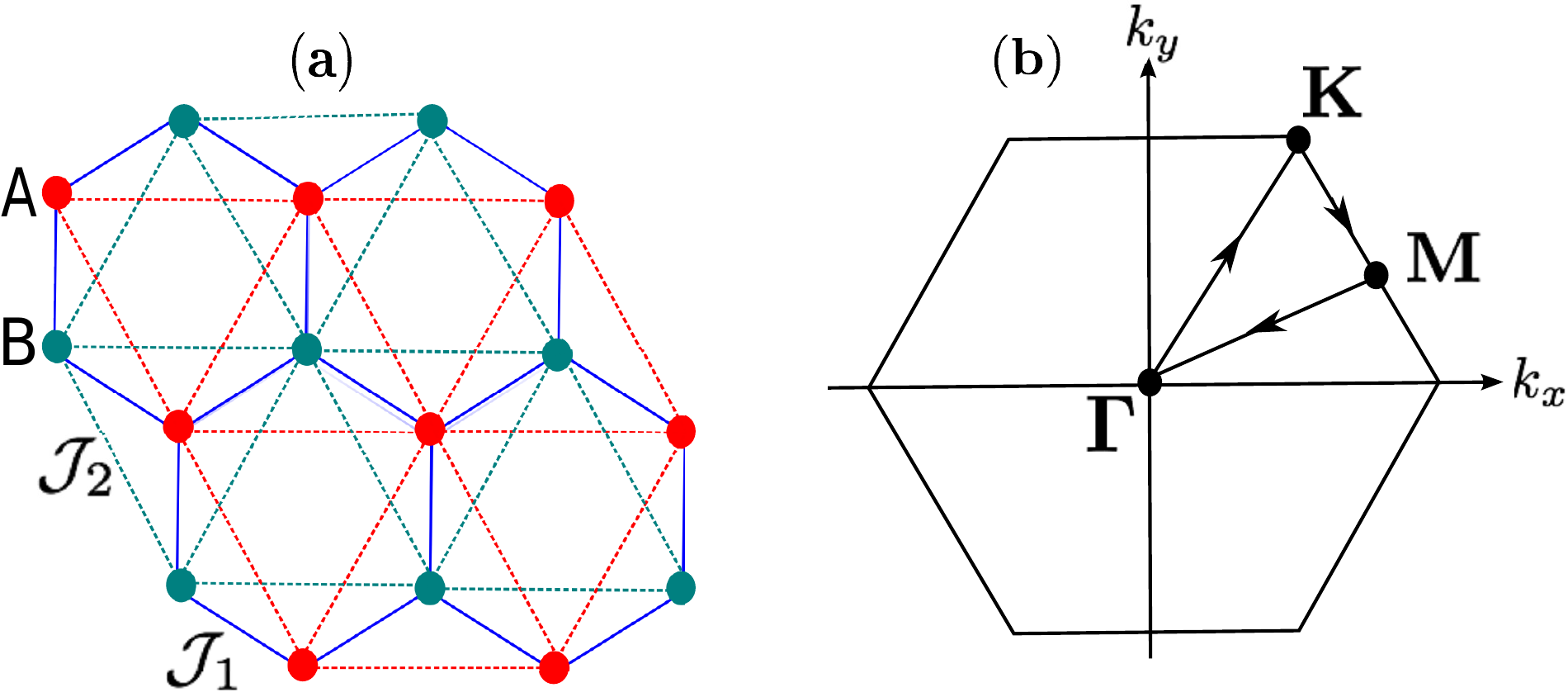}
\caption{Color online. $(a)$ Schematics of the isomorphic $\mathcal J_1$--$\mathcal J_2$  honeycomb-lattice and bilayer triangular-lattice. $(b)$ The first Brillouin zone of the system with indicated paths. }
\label{lat}
\end{figure}

The frustrated  isomorphic $\mathcal J_1$--$\mathcal J_2$ Heisenberg model on the  honeycomb- and bilayer triangular-lattice in an external magnetic field is given  by
\begin{align}
\mathcal H&=\sum_{ij}\mathcal J_{ij}\boldsymbol{\mathcal{S}}_{i}\cdot\boldsymbol{\mathcal{S}}_{j}-H\sum_{i}\mathcal S_{i}^z,
\label{model1}
\end{align}
where $\mathcal  J_{ij}=\mathcal  J_1( \mathcal J_2)$ are  nearest (next-nearest) neighbour  antiferromagnetic interactions and $H$ is the external magnetic field along the $z$-axis perpendicular to the lattice plane. The Hamiltonian \eqref{model1} has been extensively studied on the honeycomb lattice in the context of ground state (thermodynamic) properties \cite{mak0, mak1, mak2, mak3, mak4,mak6,mak7, mak8, mak9, mak10, mak11, mak12,mak13}. In the classical limit  at zero magnetic field \cite{mak0, mak1}, a collinear N\'eel order exists for $ \mathcal J_2/\mathcal J_1 < 1/6$. For $\mathcal J_2/\mathcal J_1 > 1/6$ it has a family of degenerate spiral order with  incommensurate wave vectors.  It was shown that spin wave fluctuations at leading order lift this accidental degeneracy in favour of specific wave vectors.
In particular,  the $120^\circ$ coplanar order with ordering wave vector $\bold K={\bf Q}=\lb\pm 2\pi/3\sqrt{3}, 2\pi/3\rb$ is expected to emerge for $\mathcal J_2/\mathcal J_1 \gg 1$. In this limit the Hamiltonian \eqref{model1}   is isomorphic to the bilayer TLHAF with intraplane coupling $\mathcal J_2$ and small interplane coupling $\mathcal J_1$. The two sublattices of the honeycomb lattice $A,B$ are equivalent to the top and bottom sublattices of the bilayer triangular lattice as shown in  figure \eqref{lat}. Therefore, we study both systems simultaneously via equation~\eqref{model1}.
 
 \section{Results}
 \subsection{Magnon band structures}

 It is advantageous to introduce the standard  Holstein-Primakoff  bosonization. The calculation is tedious but doable as shown in Appendix~\eqref{appena}.  We have checked that for $\mathcal J_2/\mathcal J_1 <1/6$ or equivalently $\mathcal J_2/\mathcal J_1\to 0$  the Hamiltonian recovers  the magnon band structures of collinear (canted) N\'eel antiferromagnet at $H=0$ ($H< H_s$) as well as collinear ferromagnet at $H=H_s$, where $H_s=3(2\mathcal J_1+3\mathcal J_2)$ is the saturation field. These limiting cases require the DMI for topological features to exist as previously shown \cite{sol,sol1,kkim,sol3,ran, kov}.

 \begin{figure}
\centering
\includegraphics[width=1\linewidth]{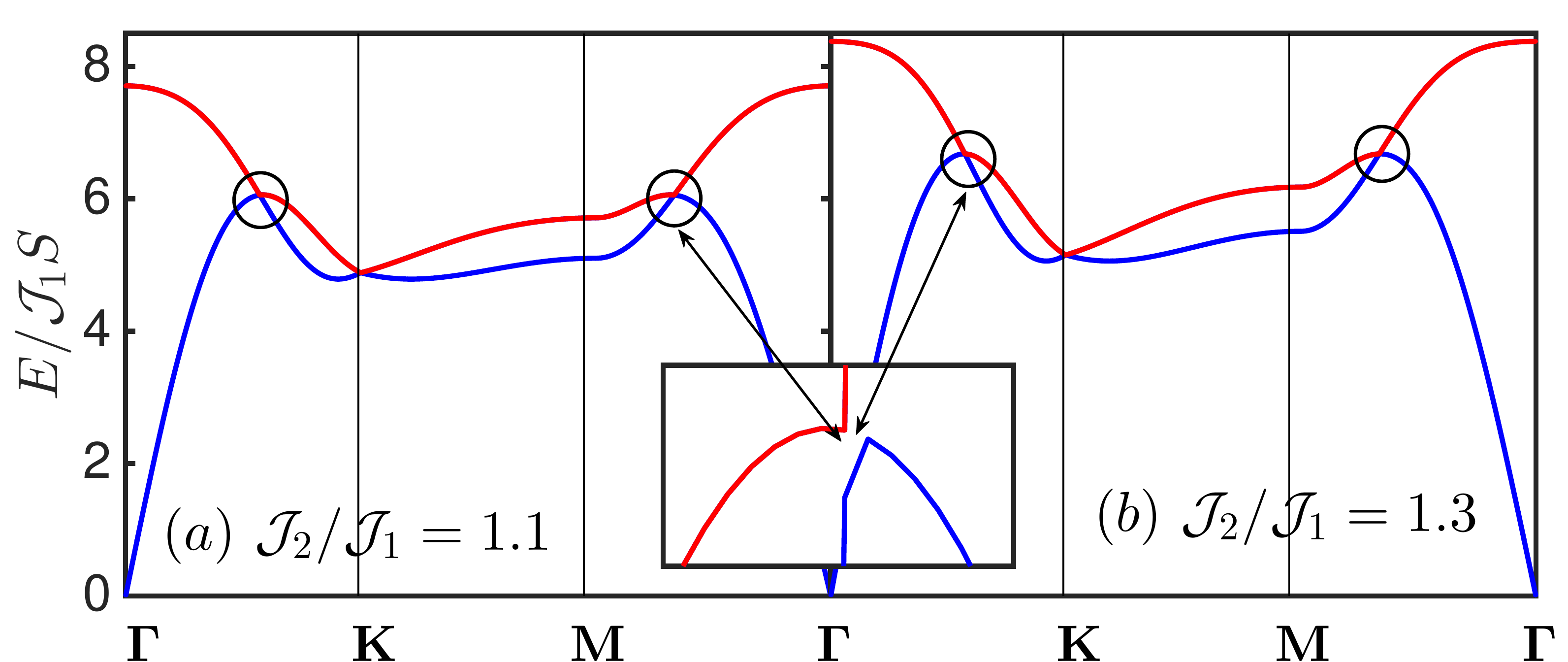}
\caption{Color online. Dirac  magnon bands of the conventional 120$^\circ$ coplanar order at zero magnetic field $H=0$. The  bands linearly touching at ${\bf K}$ and form a Dirac point. Inset shows the  circled points }
\label{plot1}
\end{figure}
\begin{figure}
\centering
\includegraphics[width=1\linewidth]{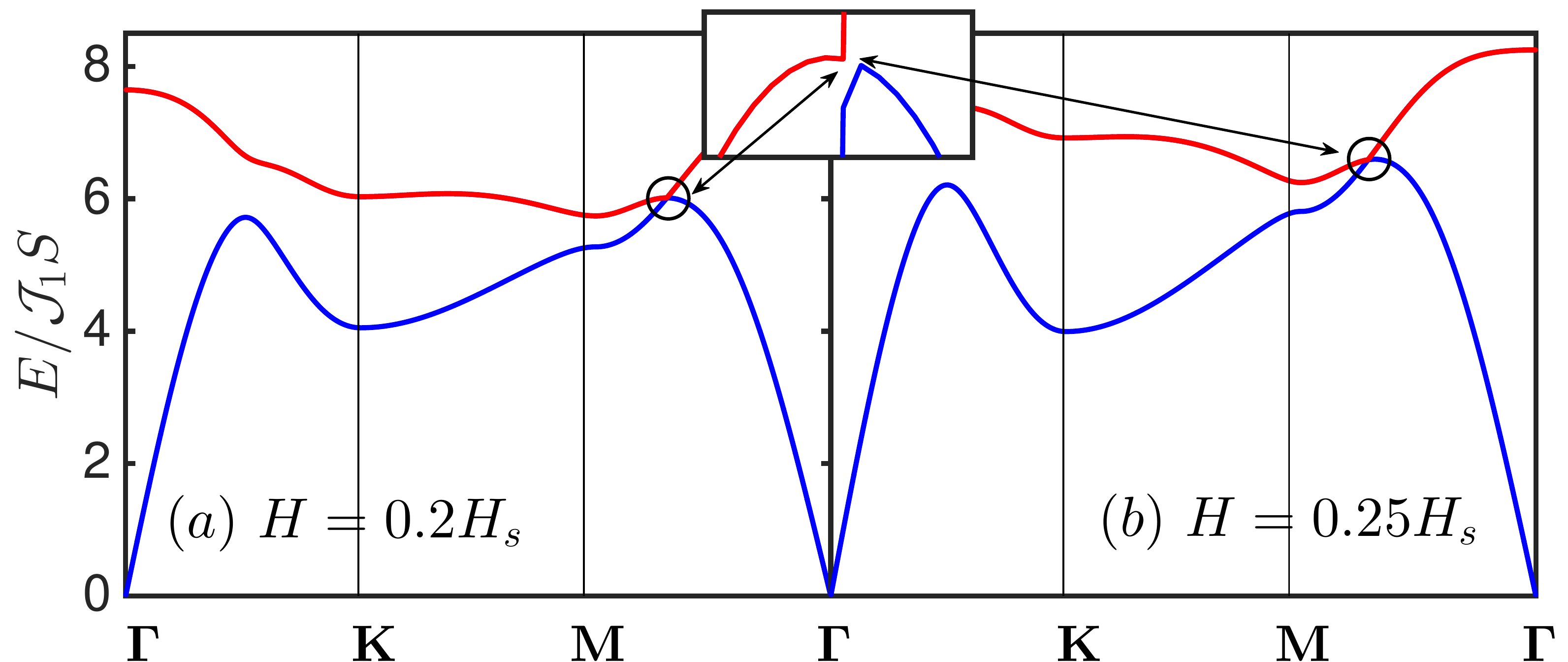}
\caption{Color online. Topological magnon bands of the non-coplanar (umbrella) spin structure for $\mathcal J_2/\mathcal J_1=1.3$ at two magnetic field values. Inset shows the  circled points.}
\label{plot2}
\end{figure}

We are interested in the dominant $\mathcal J_2$ limit corresponding to the isomorphic HLHAF and bilayer TLHAF with a stable 120$^\circ$ coplanar  order. In this regime the magnetic field induces a non-coplanar (umbrella) chiral spin texture with nonzero scalar spin chirality  $\chi$. We have shown the magnon bands at zero magnetic field $H=0$  in figure \eqref{plot1} ({\it i.e.}, conventional 120$^\circ$ spin structure). They have Dirac point nodes at ${\bf K}$. The Dirac nodes remain intact  even in the presence of (out-of-plane) DMI as it plays a stability role in certain frustrated magnets rather than a topological role.  In figure~\eqref{plot2} we have shown the magnon bands for $\mathcal J_2/\mathcal J_1=1.3$ at two values of nonzero magnetic fields.  We see that the Dirac  magnon nodes are  gapped and the magnon bands become topological due to the presence of nonzero scalar spin chirality $\chi$. Notice that there is a roton-minimum near the ordering wave vector of the coplanar  spin structure at $\bold K={\bf Q}$, which becomes gapless for large $\mathcal J_2/\mathcal J_1$. These features can be reproduced in the bilayer triangular-lattice antiferromagnetic systems as shown explicitly in Appendix~\eqref{appenb}.  
 \subsection{Berry curvature and Chern number}
   
   Most importantly, the magnon bands with nonzero scalar spin chirality $\chi$ now acquire a nonzero Berry curvature, given by 
  
  \begin{align}
\Omega_{ij;\bo s}=-\sum_{s\neq s^\prime}\frac{2\text{Im}[ \braket{\mathcal{P}_{\bo s}|v_i|\mathcal{P}_{\bo s^\prime}}\braket{\mathcal{P}_{\bo s^\prime}|v_j|\mathcal{P}_{\bo s}}]}{\lb E_{\bo s}-E_{\bo s^\prime}\rb^2},
\label{chern2}
\end{align}
where   $v_{i}=\partial (\eta\mathcal{H}_\bo)/\partial k_{i}$ defines the velocity operators and $\eta=\text{diag}(I_{N\times N},-I_{N\times N})$ is the diagonal of $N\times N$ identity matrix and $s$ labels the bands. Here $\mathcal{P}_{\bo s}$ is the paraunitary operator that diagonalizes $\eta\mathcal{H}_\bo$. The Chern numbers are given by
\begin{equation}
\mathcal{C}_s= \frac{1}{2\pi}\int_{{BZ}} dk_idk_j~ \Omega_{ij; \bo s}.
\label{chenn}
\end{equation} 
We have computed the Chern numbers numerically and established that for the two positive magnon bands, say $s=\pm$,  $\mathcal{C}_\pm=\pm \text{sgn}(\sin \phi)$, where $\sin \phi$    is related to the field-induced  scalar spin chirality $\chi$ and $\phi$ is the angle subtended by three non-coplanar (umbrella) spins. It changes sign by reversing the sign of the magnetic field or the scalar spin chirality, {\it i.e.},  $\sin \phi\to-\sin \phi$ as $\vartheta\to \vartheta +\pi$. 
\begin{figure}
\centering
\includegraphics[width=1\linewidth]{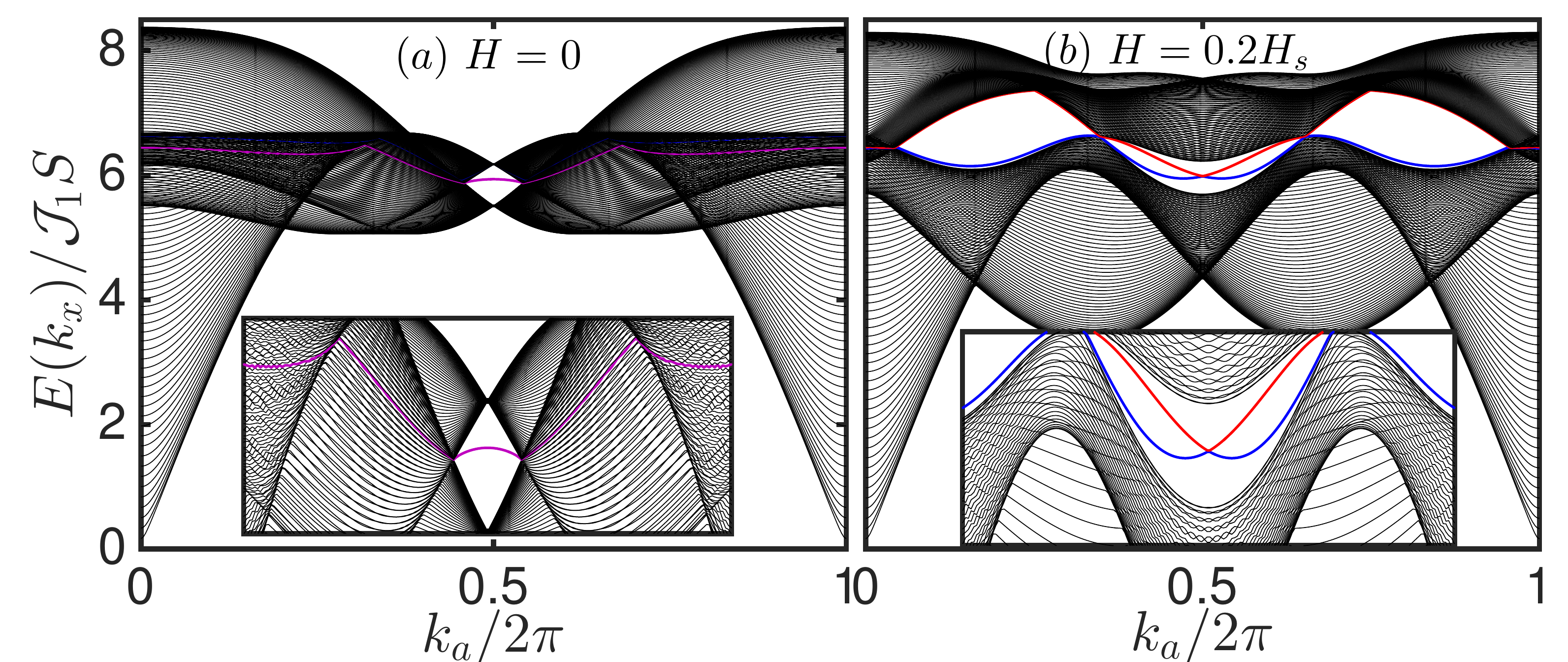}
\caption{Color online. Magnon chiral edge modes for a strip geometry periodic in $x$, but with open boundaries along $y$ for $\mathcal J_2/\mathcal J_1=1.3$.  The black region corresponds to the magnon bulk bands. The chiral edge modes are shown in colors. $(a)$ Dirac cones connected by flat chiral edge modes at zero magnetic field $H=0$. $(b)$ Gapped Dirac cones with gapless chiral edge modes for small magnetic field $H=0.2 H_s$ as a result of the emergent scalar spin chirality $\chi$. Insets show the magnified  chiral edge modes.}
\label{edge}
\end{figure}

 The existence of chiral magnon edge modes is another aspect of topological character of nontrivial magnons in insulating quantum magnets. At zero magnetic field the Dirac  magnon bulk bands are connected by a flat chiral edge mode as shown in figure~\eqref{edge}(a). As the magnetic field is turned on the flat edge modes are lifted due the presence of scalar spin chirality $\chi$ as shown in figure~\eqref{edge}(b). The chiral edge modes are now topologically protected by the Chern numbers $\mathcal{C}_\pm=\pm \text{sgn}(\sin \phi)$.
\begin{figure}
\centering
\includegraphics[width=1\linewidth]{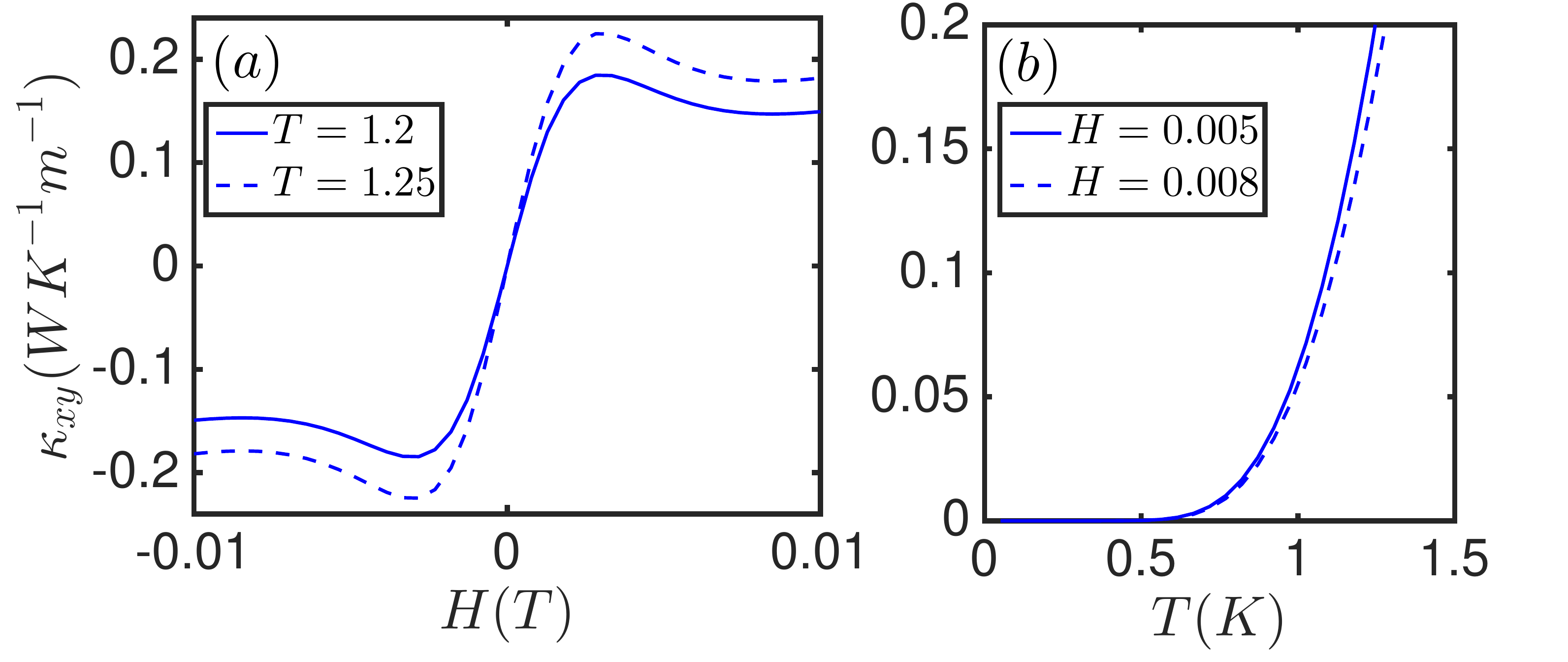}
\caption{Color online. $(a)$ Tunable thermal Hall conductivity $\kappa_{xy}$ as a function of magnetic field $H$ for two temperature values. $(b)$ Tunable  thermal Hall conductivity $\kappa_{xy}$ as a function of temperature $T$ for two magnetic field values. The coupling is set to $\mathcal J_2/\mathcal J_1=1.3$. }
\label{thm}
\end{figure}

 \subsection{Topological thermal Hall effect }

Conventionally, thermal Hall effect  is induced by the DMI as reported in  unfrustrated magnets \cite{alex0, alex2, zhh, alex4, alex4h,shin1,shin, kov1,rol,alex6,rc,alex1,alex1a,sol,sol1,kkim,ran,kov, sol3} as well as frustrated magnets \cite{rom,rom1, lau}.  For frustrated magnets with noncollinear and coplanar  spin structure thermal Hall conductivity can be nonzero in the absence of the DMI. This is possible because an external magnetic field can induce  non-coplanar spin configurations with nonzero scalar spin chirality $\chi$.  The interesting feature in the current study is that  the  coplanar  spin structure can be present in bilayer TLHAF without an intrinsic DMI  \cite{shi,ma, maksi, shi1,doi, zhou,qui}.  The thermal Hall conductivity $\kappa_{xy}$ can be derived from linear response theory \cite{shin1}.  We have shown the trends of $\kappa_{xy}$ as functions of the magnetic field and temperature  in figures~\eqref{thm} (a) and (b) respectively for a specific value of $\mathcal J_2/\mathcal J_1$. Evidently, we capture a sign change in $\kappa_{xy}$  as  the scalar spin chirality is reversed by reversing the sign of the magnetic field. Interestingly, the trend of $\kappa_{xy}$ is synthetic and tunable by the magnetic field as it  requires no intrinsic DMI. As we discuss below certain (honeycomb) triangular-lattice antiferromagnetic materials do not have an intrinsic DMI, so we do expect that $\kappa_{xy}$ can be tuned by an external magnetic field.

\section{Conclusion}

We have predicted that topological magnon bands and unconventional topological thermal Hall effect will exist in the isomorphic $\mathcal J_1$--$\mathcal J_2$  honeycomb-lattice and bilayer triangular-lattice antiferromagnets. These interesting features originate from the magnetic-field-induced  non-coplanar (umbrella) spin structure with nonzero scalar spin chirality and require no DMI in contrast to previously studied  unfrustrated magnets \cite{alex0, alex2, zhh, alex4, alex4h,shin1,shin, kov1,rol,alex6,rc,alex1,alex1a,sol,sol1,kkim,ran,kov, sol3}  and frustrated magnets \cite{rom,mcc, rom1, lau}. Therefore they are unconventional and synthetic as they can be tuned by the external magnetic field. 

Most importantly,  topological magnon features have not been previously predicted on the triangular lattice for the reasons we mentioned above. In  realistic materials the bilayer triangular antiferromagnetic systems do not usually allow an intrinsic DM  interaction,  but  an easy-plane (axis) anisotropy can be present.  This is the case in the bilayer triangular-lattice quantum antiferromagnets Ba$_3$XSb$_2$O$_9$ (X $\equiv$ Mn and Co)  with a stable 120$^\circ$ coplanar order  and no intrinsic DM  interaction \cite{doi,shi,ma, maksi, shi1, zhou,qui}. They are also quasi-two-dimensional (quasi-2D) with  dominant intraplane ($\mathcal J_2$) coupling and small interplane coupling ($\mathcal J_1$) and easy-axis anisotropy ($\Delta$). 

 The compounds VX$_2$  (X $\equiv$ Cl, Br, and I)  also form quasi-2D layered triangular-lattice quantum antiferromagnets  with a stable 120$^\circ$ coplanar order and no intrinsic DMI \cite{hir,hir1,coll}.  The net magnetic-field-induced scalar spin chirality in the non-coplanar regime will be nonzero. Therefore, the current predictions can be tested in these materials. Indeed, a finite $\kappa_{xy}$ at various external magnetic fields signifies that the magnetic excitations are topologically nontrivial. As mentioned above,   we have previously shown slightly similar results on the kagom\'e and star  lattices  \cite{sa,sa1}, but these lattice geometries naturally allow a DMI which stabilizes the coplanar spin structure \cite{el}. In contrast, the current results are different in that the bilayer triangular antiferromagnetic materials mentioned above do not have an intrinsic DMI and a non-vanishing spin-chirality can be induced by applying an external magnetic field in the plane perpendicular to the magnets.

The bilayer honeycomb-lattice quantum antiferromagnets Bi$_3$X$_4$O$_{12}$(NO$_3$) (X $\equiv$ Mn, V, and Cr) are also  promising candidates;  however it is believed that $\mathcal J_2/\mathcal J_1\ll 1$ for X $\equiv$ Mn, but an external magnetic field induces a transition to  a 3D collinear N\'eel order at $H\sim 6 T$ \cite{matt}. In the collinear regime the DMI is mandatory for topological magnons to exist, and it can be allowed in this compound \cite{oku}. There is also a possibility to synthesize different honeycomb materials with dominant $\mathcal J_2$ \cite{okk}.   We also expect the spontaneously-induced spin chirality in the chiral spin liquid  to have the same topological effects on the underlying magnetic excitations. Hence, the scalar-chirality mechanism  can help explain the recently observed thermal Hall conductivity in a spin liquid material at nonzero magnetic field \cite{wata}.
\section*{Acknowledgements}

 Research at Perimeter Institute is supported by the Government of Canada through Industry Canada and by the Province of Ontario through the Ministry of Research
and Innovation. 

\appendix

\section{Dominant $\mathcal J_2$ limit }
\label{appena}
 We are interested in the dominant $\mathcal J_2$ limit of equation \eqref{model1},  corresponding to  the isomorphic honeycomb-lattice and bilayer triangular-lattice antiferromagnets with a stable 120$^\circ$ coplanar  order.  At zero field  we take the spins to lie on the plane of the honeycomb (bilayer) triangle lattice taken as the $xy$ plane. Then, we perform a rotation  about the $z$-axis on the sublattices by the spin oriented  angles $\theta_i$.  As the external magnetic field is turned on, the spins will cant towards the direction of the field and form a non-coplanar configuration. Thus, we have to align them along the new quantization axis by performing a rotation about the $y$-axis by the field canting angle $\vartheta$. The total transformation of the spins is 
\bea 
\boldsymbol{\mathcal S}_i=\mathcal{R}_z(\theta_i)\cdot\mathcal{R}_y(\vartheta)\cdot\boldsymbol{\mathcal S}_i^\prime,
\label{trans}
\eea
where
\begin{align}
\mathcal{R}_z(\theta_i)\cdot\mathcal{R}_y(\vartheta)
=\begin{pmatrix}
\cos\theta_i\cos\vartheta & -\sin\theta_i & \cos\theta_i\sin\vartheta\\
\sin\theta_i\cos\vartheta & \cos\theta_i &\sin\theta_i\sin\vartheta\\
-\sin\vartheta & 0 &\cos\vartheta
\end{pmatrix}.
\label{eq3}
\end{align}
Next, we plug the spin transformation \eqref{trans} into the Hamiltonian \eqref{model1}. There are numerous terms but we retain only the terms that contribute to the free magnon model,  given by

  \begin{align}
  \mathcal H&= \sum_{ ij} \mathcal J_{ij}\big[\cos\theta_{ij} \boldsymbol{\mathcal S}_{i}^\prime\cdot \boldsymbol{\mathcal S}_{j}^\prime+ \sin\theta_{ij}\cos\vartheta \hat{\bold z}\cdot\lb \boldsymbol{\mathcal S}_{i}^\prime\times\boldsymbol{\mathcal S}_{j}^\prime\rb  \\&\nonumber +2\sin^2\lb\frac{\theta_{ij}}{2}\rb\lb\sin^2\vartheta \mathcal S_{i}^{\prime x}\mathcal S_{j}^{\prime x} +\cos^2\vartheta \mathcal S_{i}^{\prime z}\mathcal S_{j}^{\prime z}\rb\big] \\&\nonumber-H\cos\vartheta\sum_{i} \mathcal S_{i}^{\prime z},
  \end{align}
where $\theta_{ij}=\theta_i-\theta_j$. We note that for $\mathcal J_{ij}=\mathcal J_1$, $\sin\theta_{ij}=0$, therefore  the field-induced scalar spin chirality of the non-coplanar  (umbrella) spin configurations defined as $\chi= \sum \boldsymbol{\mathcal S}_i^\prime\cdot\lb \boldsymbol{\mathcal S}_{j}^\prime\times\boldsymbol{\mathcal S}_{k}^\prime\rb$  is induced only within the triangular plaquettes of the NNN bonds on the honeycomb lattice or the triangular plaquettes of the NN bilayer triangular lattice. Here $\sin\theta_{ij}=\nu_{ij}|\sin\theta_{ij}|$,  where $\nu_{ij}=\pm 1$ denotes the sign of the magnon hopping along the triangular plaquettes of the honeycomb (bilayer triangular) lattice.
  
  Usually, the net chirality vanishes on the triangular lattice because neighbouring triangular plaquettes contribute equal and opposite chirality \cite{alex0}. But the field-induced spin chirality of the non-coplanar  (umbrella)  spin structure will be finite as it is coupled to the magnetization of the non-coplanar spin configuration. The sign of the scalar-chirality is determined by the magnetic field and it has the same sign on each honeycomb (bilayer) triangle for $H>0$, whereas for $H<0$ the spins on each honeycomb (bilayer) triangle flip, now $\vartheta \to \pi+\vartheta $ on each triangle. In this case the net scalar-chirality is nonzero \cite{gro}.
  The origin of the spin chirality can also be inferred from geometric frustration of the lattice, which can allow a chiral spin liquid phase.   In this case, the  scalar spin chirality can be spontaneously developed.  Because of the scalar spin chirality the system has already acquired a real space Berry curvature from the chiral magnetic spin structure.
  We therefore expect the spontaneously-induced and the field-induced spin chirality to have the same topological effects on the underlying magnetic excitations.

In the present case, it is advantageous to introduce the  Holstein-Primakoff  bosonization \cite{hp}: $
 \mathcal S_{i}^{z}= S-a_{i}^\dagger a_{i},~ \mathcal S_{i}^{+} \approx  \sqrt{2S}a_{i}=\lb\mathcal S_{i}^{-}\rb^\dg$, where $\mathcal S_{i}^{\pm}=\mathcal S_{i}^{x}\pm i \mathcal S_{i}^{y}$ and $a_{i}^\dagger(a_{i})$ are the bosonic creation (annihilation) operators. 
 The magnon tight binding Hamiltonian is given by
\begin{align}
\mathcal H_{\mathcal J_1}&= S\sum_{\la i,j\ra}\big[ t_{1,z}(a_i^\dg a_i +a_j^\dg a_j) + t_{1,r}(a_i^\dg a_j + h.c.)\\&\nonumber + t_{1,o}(a_i^\dg a_j^\dg + h.c.)\big],
\end{align}
\begin{align}
\mathcal H_{\mathcal J_2}&=  S\sum_{\la\la i,j\ra\ra}\big[t_{2,z}(a_i^\dg a_i +a_j^\dg a_j) +t_{2} (e^{-i\phi_{ij}}a_i^\dg a_j + h.c.)\\&\nonumber +t_{2,o}(a_i^\dg a_j^\dg + h.c.)\big] ;~\mathcal H_{H}=H_{\vartheta}\sum_{i}a_i^\dg a_i,
\end{align}
where $\la i,j\ra$ and $\la\la i,j\ra\ra$ denote the summations over the NN and NNN sites respectively.
\begin{align}
& t_{1,z}=\mathcal J_{1}[1-2\cos^2\vartheta)],~
t_{1,r}=- \mathcal J_{1}\big[1 -\sin^2\vartheta\big],\\
&t_{1,o}=\mathcal J_{1} \sin^2\vartheta,~
t_{2,z}= \frac{\mathcal J_{2}}{2}[1-3\cos^2\vartheta],\\
&t_2=\sqrt{(t_{2,r})^2+(t_{2,m})^2},~
t_{2,r}= -\frac{\mathcal J_{2}}{2}(1 -3\sin^2\vartheta/2),\\
&t_{2,m}= \frac{\sqrt 3\mathcal J_{2}}{2}\cos\vartheta,~
t_{2,o}=\frac{3\mathcal J_{2}}{4} \sin^2\vartheta,
\end{align}
and $H_\vartheta=H\cos\vartheta$.  The angle $\vartheta$ is determined from the mean-field energy, given by
\begin{align}
E_0&= -\frac{3\mathcal J_1}{2}\lb 1 - 2\cos^2\vartheta\rb-\frac{3}{2}\mathcal J_2\lb 1 - 3\cos^2\vartheta\rb-H\cos\vartheta,
\end{align}
where $E_0=E_{MF}/NS^2$ and $N$ is the total number of sites on the honeycomb lattice.  The magnetic field is rescaled in unit of $S$. Minimizing this energy yields the canting angle $\cos\vartheta = H/H_s$, where $H_s=3(2\mathcal J_1+3\mathcal J_2)$ is the saturation field. The  solid angle subtended by three non-coplanar spins  is given by $\phi_{ij}=\nu_{ij}\phi$, where $\phi=\tan^{-1}[t_{2,m}/t_{2,r}]$. In Fourier space the Hamiltonian can be written as $\mathcal H=\frac{1}{2}S\sum_{\bo} \Psi_\bo^\dg\mathcal{H}_{\bo}\Psi_\bo +\text{const.}$, where $\Psi_\bo=(\psi_\bo^\dg,\psi_{-\bo})$, with $\psi_\bo^\dg=(a_{\bo,A}^\dg,a_{\bo,B}^\dg)$.

\begin{align}
\mathcal{H}_{\bo}=
\begin{pmatrix}
I_{\bo}-m_\bo &t_{1,r} f_{\bo}^* &t_{2,o} \lambda_{\bo}^*&t_{1,o}f_{\bo}^*\\
t_{1,r} f_{\bo}&I_{\bo} +m_\bo &t_{1,o}f_{\bo}&t_{2,o} \lambda_{\bo}^*\\
t_{2,o} \lambda_{\bo}&t_{1,o}f_{\bo}^*&I_{\bo} +m_\bo& t_{1,r} f_{\bo}^* \\
t_{1,o}f_{\bo}&t_{2,o} \lambda_{\bo}& t_{1,r} f_{\bo}&I_{\bo} -m_\bo
\end{pmatrix},
\label{outp}
\end{align}
where $f_{\bo}= 1 + e^{-ik_a}+e^{-i(k_a +k_b)}$, $\lambda_{\bo}=2[\cos k_a +\cos k_b+\cos(k_a+k_b)]$;  $m_\bo=2t_2 \sin\phi [\sin k_a +\sin k_b-\sin(k_a+k_b)]$;  $I_{\bo}=3  t_{1,z} +6 t_{2,z}+t_2\cos\phi\lambda_{\bo} +H\cos\vartheta=3(\mathcal J_1 +\mathcal J_2)+ t_2\cos\phi\lambda_{\bo} $. The vectors are  $\hat a=\sqrt{3}\hat x$ and $\hat b= -\sqrt{3}\hat x/2 + 3\hat y/2$ with $k_a=\bo\cdot\hat a$ and $k_b=\bo\cdot\hat b$. 
 We diagonalize the Hamiltonian numerically via the generalized Bogoluibov transformation \cite{bla}. For $\mathcal J_2/\mathcal J_1 <1/6$ or equivalently $\mathcal J_2/\mathcal J_1\to 0$  the Hamiltonian recovers  collinear (canted) N\'eel antiferromagnet at $H=0$ ($H< H_s$) as well as collinear ferromagnet at $H=H_s$. These limiting cases require the DM interaction for topological features to exist as previously shown \cite{sol,sol1,kkim,sol3,ran, kov}. The dominant $\mathcal J_2$ limit is different and requires no DM interaction for topological features to exist.
\begin{figure}
\centering
\includegraphics[width=3in]{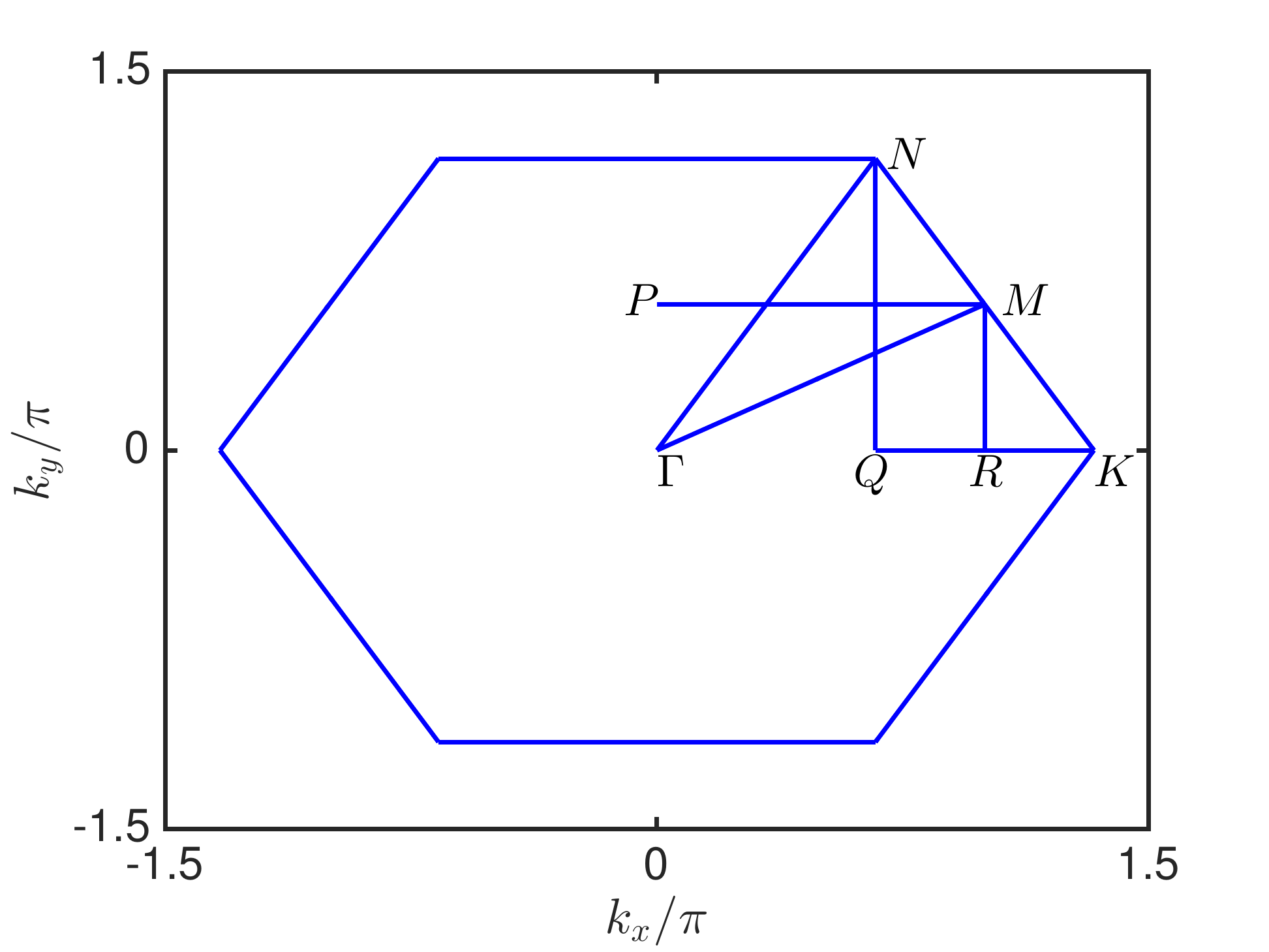}
\caption{Color online.  The first Brillouin zone of the triangular lattice and the corresponding paths that will be adopted in this section.}
\label{bb}
\end{figure} 

 \section{Bilayer triangular-lattice antiferromagnets} 
 \label{appenb}
The dominant $\mathcal J_2$ limit of frustrated honeycomb lattice is isomorphic to the bilayer triangular-lattice Heisenberg antiferromagnets.  The conventional Hamiltonian for bilayer triangular-lattice Heisenberg antiferromagnets  applicable to real materials is given  by
 \begin{align}
\mathcal H&=\mathcal J_2\sum_{\la i,j\ra,\tau}\big[\boldsymbol{\mathcal{S}}_{i\tau}^\perp\cdot\boldsymbol{\mathcal{S}}_{j\tau}^\perp +\Delta\mathcal S_{i\tau}^z\mathcal S_{j\tau}^z\big]-H\sum_{i,\tau}\mathcal S_{i,\tau}^z\label{tri}\\&\nonumber +\mathcal J_1\sum_{\la i,j\ra,\tau\tau^\prime}\big[\boldsymbol{\mathcal{S}}_{i\tau}^\perp\cdot\boldsymbol{\mathcal{S}}_{j\tau^\prime}^\perp +\Delta\mathcal S_{i\tau}^z\mathcal S_{j\tau^\prime}^z\big],
\end{align}
where $\tau$ labels the top and bottom layers and $\boldsymbol{\mathcal{S}}_{i}^\perp=(\mathcal{S}_{i}^x, \mathcal{S}_{i}^y)$.  Note that all the interactions are now nearest-neighbour (NN). The easy-plane anisotropy lies in the range $0\leq \Delta\leq 1$.  The NN intraplane coupling is $\mathcal J_2>0$ and the NN interplane coupling is $\mathcal J_1>0$ with $\mathcal J_1\ll \mathcal J_2$, i.e., quasi-2D limit.
After the  rotation  in the spin space \eqref{trans} and \eqref{eq3} we have
 \begin{align}
  \mathcal H&= \sum_{ ij} \mathcal J_{ij}\big[\cos\theta_{ij} \boldsymbol{\mathcal S}_{i}^\prime\cdot \boldsymbol{\mathcal S}_{j}^\prime+ \sin\theta_{ij}\cos\vartheta \hat{\bold z}\cdot\lb \boldsymbol{\mathcal S}_{i}^\prime\times\boldsymbol{\mathcal S}_{j}^\prime\rb  \\&\nonumber +(\Delta -\cos\theta_{ij})\lb\sin^2\vartheta \mathcal S_{i}^{\prime x}\mathcal S_{j}^{\prime x} +\cos^2\vartheta \mathcal S_{i}^{\prime z}\mathcal S_{j}^{\prime z}\rb\big] \\&\nonumber-H\cos\vartheta\sum_{i} \mathcal S_{i}^{\prime z},
  \end{align}
where we have retained the free magnon model. $\mathcal  J_{ij}=\mathcal  J_1( \mathcal J_2)$. Because of the antiferromagnetic interplane coupling the spins on the top layer are orientated in the opposite direction to those on the bottom layer, hence $\sin\theta_{ij}=0$ for $\mathcal  J_{ij}=\mathcal  J_1$ and the scalar-chirality vanishes. However, each layer form a $120^\circ$ coplanar order and $\sin\theta_{ij}=\nu_{ij}\sin(120^\circ)$ for  $\mathcal  J_{ij}=\mathcal  J_2$, where $\nu_{ij}=\pm$ for magnon hopping on the top and bottom layers respectively.  The scalar-chirality of the non-coplanar structure on both layers are along the positive $z$-axis for $H>0$,  whereas for $H<0$ the spins on each triangular-layer flip by $180^\circ$. Now $\vartheta \to \pi+\vartheta $  on each layer. Therefore the net scalar-chirality is nonzero in both cases \cite{gro}.

\begin{figure}
\centering
\includegraphics[width=1\linewidth]{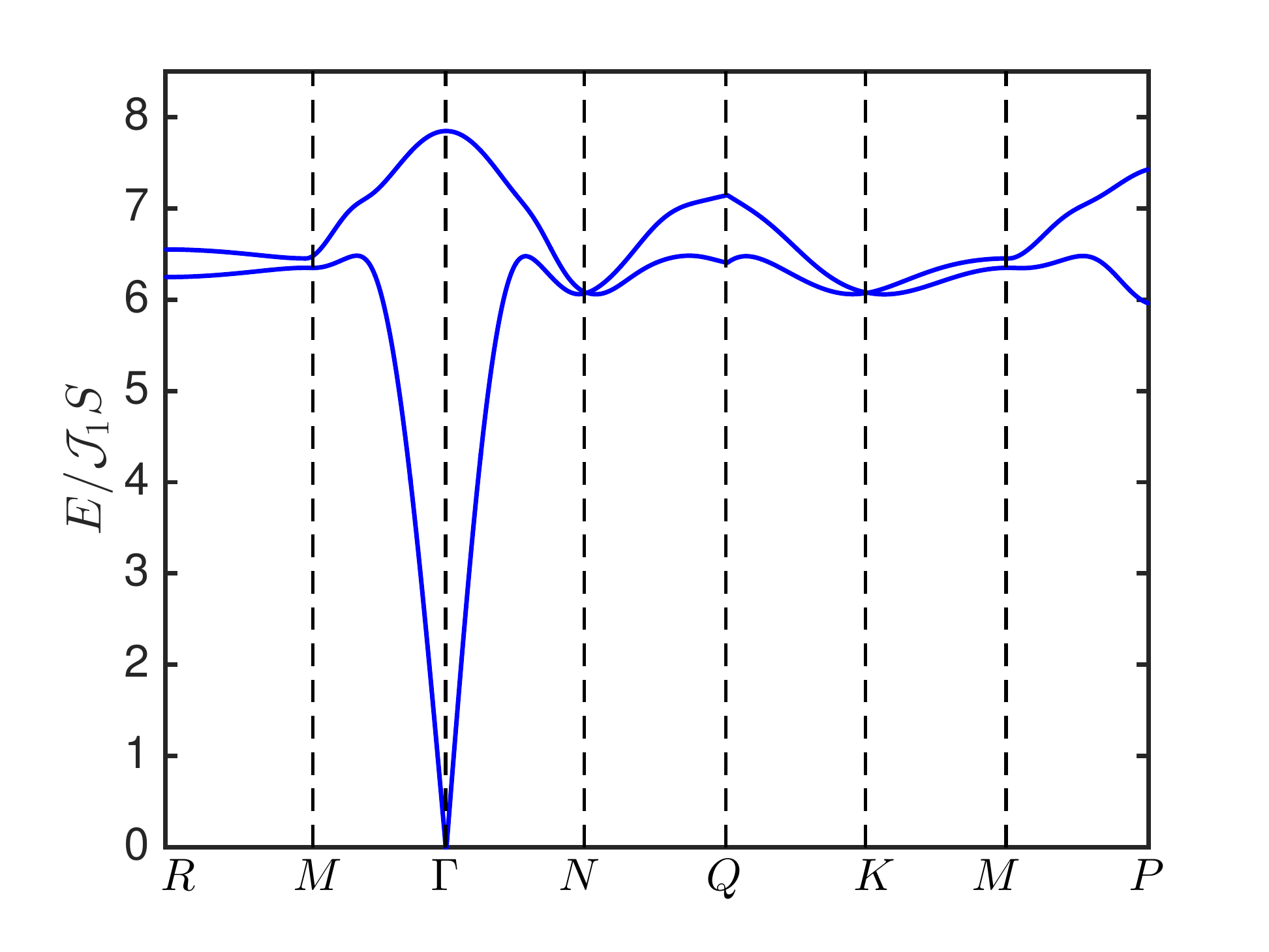}
\caption{Color online. Dirac magnon bands (at $N$ and $K$) of the bilayer XXZ triangular-lattice  antiferromagnet with the conventional 120$^\circ$ coplanar order at zero magnetic field $H=0$. The plot is generated with $\Delta=0.7$ and $\mathcal J_2/\mathcal J_1=1.3$.}
\label{sm1}
\end{figure}
\begin{figure}
\centering
\includegraphics[width=1\linewidth]{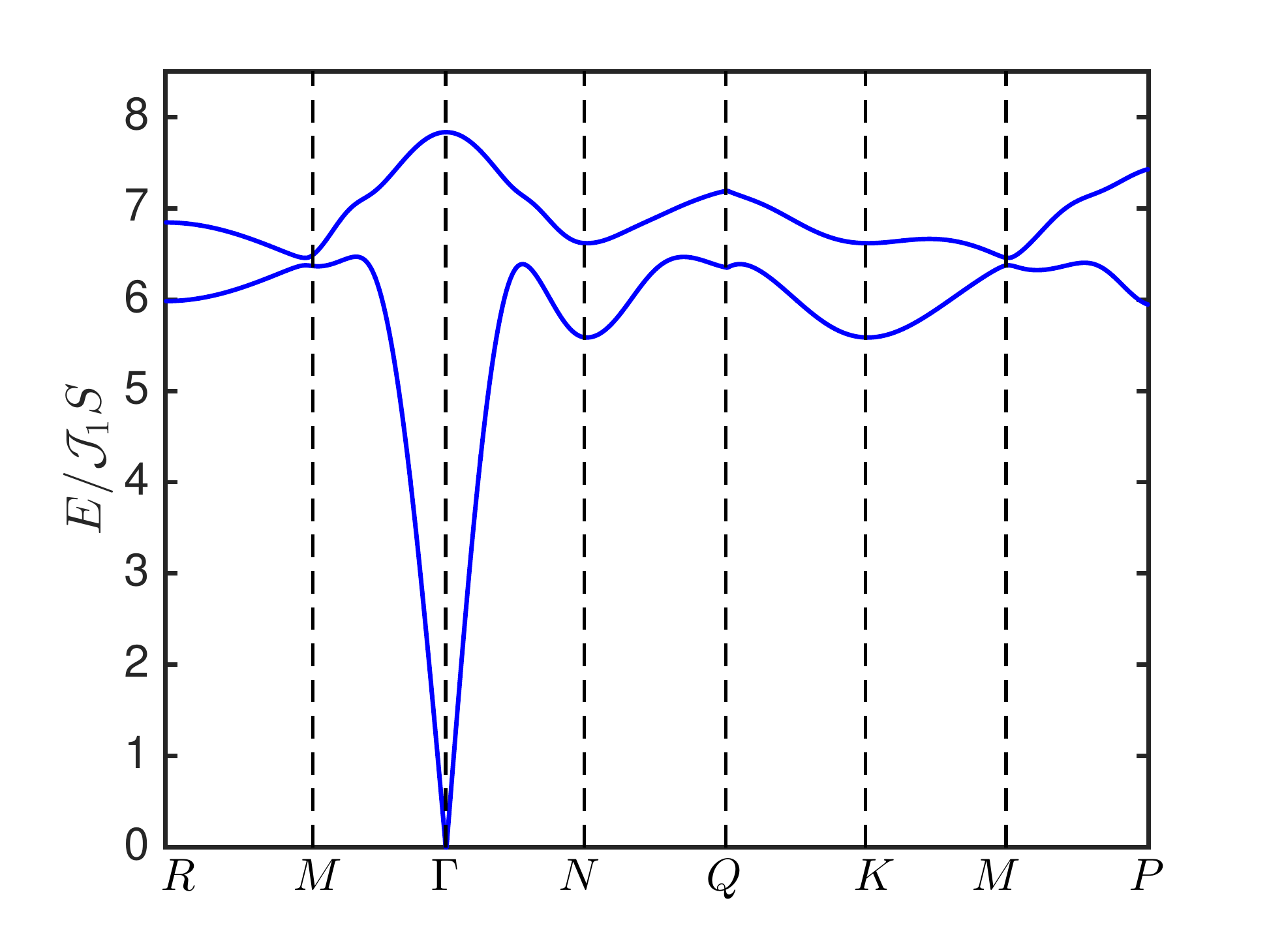}
\caption{Color online. Topological magnon bands of the bilayer XXZ triangular-lattice  antiferromagnet with the non-coplanar (umbrella) spin configuration at nonzero magnetic field $H=0.2H_s$. The plot is generated with $\Delta=0.7$ and $\mathcal J_2/\mathcal J_1=1.3$.}
\label{sm2}
\end{figure}

We adopt the one-sublattice structure \cite{sol4,sol5}  on each layer  of the triangular lattice and label them $A$ and $B$. From figure 1 (a) in the main text, we see that the stacking of the bilayer triangle is such that there are six nearest-neighbours on each layer and three nearest neighbours between the layers.  The parameters of the corresponding tight-binding model are
   
\begin{align}
& t_{1,z}=\mathcal J_{1}[1-(1+\Delta)\cos^2\vartheta)],~
t_{1,r}=- \mathcal J_{1}\big[1 -\frac{1+\Delta}{2}\sin^2\vartheta\big],\\
&t_{1,o}=\frac{\mathcal J_{1}(1+\Delta)}{2} \sin^2\vartheta,~
t_{2,z}= \frac{\mathcal J_{2}}{2}[1-(2\Delta +1)\cos^2\vartheta],\\
&t_2=\sqrt{(t_{2,r})^2+(t_{2,m})^2},~
t_{2,r}= -\frac{\mathcal J_{2}}{2}\lb 1 -\frac{(2\Delta +1)}{2}\sin^2\vartheta\rb,\\
&t_{2,m}= \frac{\sqrt 3\mathcal J_{2}}{2}\cos\vartheta,~
t_{2,o}=\frac{\mathcal J_{2}(2\Delta+1)}{4} \sin^2\vartheta,
\end{align}
with $\cos\vartheta=H/H_s$ and $H_s=3\mathcal J_1(1+\Delta)+3\mathcal J_2(2\Delta +1)$. 

In the basis $\Psi_\bo^\dg=(a_{\bo,A}^\dg,a_{-\bo,A},a_{\bo,B}^\dg,a_{-\bo,B})$, the momentum space Hamiltonian is given by \begin{align}
\mathcal{H}_{\bo}=
\begin{pmatrix}
I_{\bo}-m_\bo &t_{2,o} \lambda_{\bo} &t_{1,r}f_{\bo}^*&t_{1,o}f_{\bo}^*\\
t_{2,o}\lambda_{\bo}^*&I_{\bo} +m_\bo &t_{1,o}f_{\bo}&t_{1,r}f_{\bo}\\
t_{1,r}f_{\bo} &t_{1,o}f_{\bo}^*&I_{\bo} +m_\bo& t_{2,o}\lambda_{\bo} \\
t_{1,o}f_{\bo}&t_{1,r}f_{\bo}^*& t_{2,o}\lambda_{\bo}^* &I_{\bo} -m_\bo
\end{pmatrix},
\label{outp}
\end{align}
where $f_{\bo}= 1 + e^{-ik_1}+e^{-i(k_1+k_2)}$,  $\lambda_{\bo}=2[\cos k_1 +\cos k_2+\cos (k_1+k_2)]$;  $m_\bo=2t_2 \sin\phi [\sin k_1 +\sin k_2-\sin (k_1+k_2)]$;  $I_{\bo}=  3t_{1,z} +6 t_{2,z}+t_2\cos\phi\lambda_{\bo} +H\cos\vartheta=3\mathcal J_1 +3\mathcal J_2+ t_2\cos\phi\lambda_{\bo} $. Here  $k_i=\bo\cdot\hat e_i$ and the primitive vectors of the triangular lattice are  $\hat e_1=\hat x$  and $\hat e_2= -\hat x/2 + \sqrt{3}\hat y/2$.  The Brillouin zone  paths are depicted in figure~\eqref{bb}.

It is evident that when the interplane coupling vanishes, {\it i.e.}, $\mathcal J_1 =0=t_{1,r}=t_{1,o}$, the Hamiltonian reduces to  two decoupled triangular-lattice XXZ antiferromagnets. For $\mathcal J_1\neq 0$ we have shown the magnon bands of the bilayer triangular-lattice XXZ antiferromagnets at $H=0$ and $H=0.2H_s$ respectively  in figures.~\eqref{sm1} and \eqref{sm2}. We see that the magnon bands have the same structure as the frustrated honeycomb lattice shown in the main text. Topological magnon bands in figure~\eqref{sm2} directly imply the existence of nonzero Chern numbers, magnon edge modes, and thermal Hall conductivity. We note that $\Delta=1$ has the same topological features as expected.


\begin{thebibliography}{99}
 

  \bibitem{alex6}
M. Hirschberger et al.,  Phys. Rev. Lett. {\bf 115}, 106603 (2015).
\bibitem{rc}
R. Chisnell et al., Phys. Rev. Lett. {\bf 115}, 147201  (2015).


    \bibitem{alex1}
Y. Onose et al.,  Science  { \bf 329}, 297 (2010).
 \bibitem{alex1a}
T. Ideue et al., Phys. Rev. B. {\bf 85}, 134411 (2012).
\bibitem{alex0}
 H. Katsura, N. Nagaosa, and P. A. Lee,   Phys. Rev. Lett.  {\bf 104},  066403 (2010).
  \bibitem{alex2}
 R. Matsumoto and S. Murakami, Phys. Rev. Lett. {\bf 106}, 197202 (2011).  Phys. Rev. B. {\bf 84}, 184406 (2011).
 \bibitem{zhh} 
  L. Zhang, J. Ren, J. S. Wang, and B. Li, Phys. Rev. B {\bf 87}, 144101 (2013).
  
  \bibitem{alex4}
A.  Mook, J.  Henk, and I. Mertig, Phys. Rev. B {\bf 90}, 024412 (2014).   Phys. Rev. B {\bf 89}, 134409 (2014).
\bibitem{shin}
R. Shindou et al., Phys. Rev. B 87,  174427 (2013).
   \bibitem{shin1}
R. Matsumoto, R. Shindou, and S. Murakami, Phys. Rev. B 89, 054420 (2014).

 \bibitem{alex4h} 
 H. Lee, J. H. Han, and P. A. Lee,   Phys. Rev. B.  {\bf 91},  125413 (2015).
 \bibitem{kov1}
 A.  A. Kovalev and V.  Zyuzin, Phys. Rev. B {\bf 93}, 161106(R) (2016).
  \bibitem{rol}
A. Rold\'an-Molina, A. S. Nunez, and J. Fern\'andez-Rossier,
New J. Phys. {\bf 18}, 045015 (2016).

 \bibitem{sol}
S. A.  Owerre, J. Phys.: Condens. Matter {\bf 28}, 386001 (2016). 

\bibitem{sol1}
S. A.  Owerre, J. Appl. Phys. {\bf 120}, 043903 (2016).
\bibitem{kkim}
S. K. Kim et al., Phys. Rev. Lett. 117, 227201 (2016).

 \bibitem{ran}
R. Cheng, S. Okamoto, D. Xiao, Phys. Rev. Lett. 117, 217202 (2016).
\bibitem{kov}
 V.  Zyuzin and  A.  A. Kovalev, Phys. Rev. Lett. 117, 217203 (2016).
\bibitem{sol3}
S. A.  Owerre, J. Appl. Phys. {\bf 121}, 223904 (2017).
\bibitem{dm}
 I. Dzyaloshinsky, J. Phys. Chem. Solids {\bf 4}, 241 (1958).
  \bibitem{dm2}
   T. Moriya, Phys. Rev. {\bf 120}, 91 (1960).
     \bibitem{fdm}
F. D. M. Haldane, Phys. Rev. Lett. {\bf 61}, 2015 (1988).
   \bibitem{yu3}
C. L. Kane and E.J. Mele,  Phys. Rev. Lett. {\bf 95}, 226801 (2005).

 \bibitem{sa}
  S. A. Owerre,  Phys. Rev. B {\bf 95}, 014422 (2017).
  \bibitem{sa1}
 S. A. Owerre, J. Phys.: Cond. Mat. 29, 03LT01 (2017).
   \bibitem{el}
M. Elhajal, B. Canals, and C. Lacroix, Phys. Rev. B {\bf 66},
014422 (2002).
\bibitem{matt}
M. Matsuda et al., Phys. Rev. Lett. {\bf 105}, 187201 (2010).

 \bibitem{mak0}
A. Mulder et al., Phys. Rev. B. {\bf 81}, 214419 (2010).

 \bibitem{mak1}
 R. Ganesh et al., Phys. Rev. B. {\bf 83}, 144414 (2011).

 \bibitem{mak2}
 J. Oitmaa and R. R. P. Singh,  Phys. Rev. B. {\bf 85}, 014428 (2012).
   \bibitem{mak3}
 H. Zhang, M. Arlego, and C. A. Lamas, Phys. Rev. B. {\bf 89}, 024403 (2014).
    \bibitem{mak4}
 F. A. G\'omez Albarrac\'in and H. D. Rosales,  Phys. Rev. B. {\bf 92}, 144413 (2016).
 \bibitem{mak6}
 F. Wang, Phys. Rev. B {\bf 82}, 024419 (2010).
 \bibitem{mak7}
 Y.-M. Lu and Y. Ran, Phys. Rev. B {\bf 84}, 024420 (2011).
 \bibitem{mak8}
 B. K. Clark, D. A. Abanin, and S. L. Sondhi, Phys. Rev.
Lett. 107, 087204 (2011).
  \bibitem{mak9}
H. Mosadeq, F. Shahbazi, and S.A. Jafari, J. Phys.
Condens. Matter {\bf 23}, 226006 (2011).
  \bibitem{mak10}
A. F. Albuquerque et al., Phys. Rev. B {\bf 84}, 024406
(2011).
\bibitem{mak11}
R. Bishop et al., J. Phys.
Condens. Matter {\bf 24}, 236002 (2012).
\bibitem{mak12}
F. Mezzacapo and M. Boninsegni, Phys. Rev. {\bf B} 85,
060402 (2012).
\bibitem{mak13}
R. Ganesh, J. van den Brink, and S. Nishimoto
Phys. Rev. Lett. {\bf 110}, 127203 (2013).
\bibitem{jian}
K. Jiang et al., Phys. Rev. Lett. {\bf 114}, 216402 (2015).

 \bibitem{gro}
D. Grohol et al.,  Nat. Mater. {\bf 4}, 323 (2005).
\bibitem{doi}
Y. Doi, Y. Hinatsu, and K. Ohoyama, J. Phys. Condens. Matter {\bf 16}, 8923 (2004).
\bibitem{shi}
Y. Shirata et al., Phys. Rev. Lett. {\bf 108}, 057205 (2012).
\bibitem{zhou}
H. D. Zhou et al., Phys. Rev. Lett. {\bf 109}, 267206 (2012).
\bibitem{shi1}
T. Susuki et al., Phys. Rev. Lett. {\bf 110}, 267201 (2013).
\bibitem{qui}
G. Quirion et al., Phys. Rev. B {\bf 92}, 014414 (2015).
\bibitem{maksi}
P. A. Maksimov, M. E. Zhitomirsky, and A. L. Chernyshev, Phys. Rev. B {\bf 94}, 140407(R) (2016).
\bibitem{ma}
J. Ma et al., Phys. Rev. Lett. {\bf 116}, 087201 (2016).

\bibitem{hir}
K. Hirakawa, H. Kadowaki, and K. Ubukoshi, J. Phys. Soc. Jpn. {\bf 52}, 1814 (1983).
\bibitem{hir1}
 H. Kadowaki, K. Ubukoshi, and K. Hirakawa,  J. Phys. Soc. Jpn., {\bf 54}, 363 (1985).
 \bibitem{coll}
 M. F. Collins and O. A. Petrenko, Can. J. Phys. {\bf 75} 605 (1997).
 
\bibitem{tak}
H. Takatsu et al., Phys. Rev. Lett. {\bf 105}, 137201 (2010).

\bibitem{rom}
J. Romh\'anyi, K. Penc, and R. Ganesh, Nat. Commun. {\bf 6}, 6805
(2015).
\bibitem{rom1}
M. Malki and K. P. Schmidt, Phys. Rev. B {\bf 95}, 195137 (2017).

\bibitem{lau}
P. Laurell, G. A. Fiete, 	Phys. Rev. Lett. {\bf 118}, 177201 (2017).


\bibitem{mcc}
P. A. McClarty et al., Nat. Phys., doi:10.1038/nphys4117 (2017).

 

  \bibitem{oku}
S. Okubo et al., Phys. Rev. B {\bf 86}, 140401(R) (2012).

   
\bibitem{okk}
Y. Singh and P. Gegenwart, Phys. Rev. B {\bf 82}, 064412 (2010).
\bibitem{wata}
D. Watanabe et al., Proc. Natl. Acad. Sci. USA 113, 8653 (2016).
 \bibitem{bla}
J. -P. Blaizot and G. Ripka, {\it Quantum Theory of Finite
Systems} (MIT Press, Cambridge, MA, 1986).
\bibitem{hp}
T. Holstein and H. Primakoff, Phys. Rev. {\bf 58}, 1098 (1940).


\bibitem{sol4}
A. V. Chubukov et al., J. Phys.: Condens. Matter {\bf 6}, 8891 (1994).
\bibitem{sol5}
S. Sachdev, Phys. Rev. B {\bf 45}, 12377 (1992).
\end{thebibliography}
\end{document}